\documentclass[reprint,aps,pra,twocolumn,showpacs,superscriptaddress]{revtex4-1}  % for review and submission
\usepackage{graphicx}  % needed for figures
\usepackage{dcolumn}   % needed for some tables
\usepackage{bm}        % for math
\usepackage{amssymb}   % for math
\usepackage{amsmath}
\usepackage[german,english]{babel}
\usepackage[T1]{fontenc}
\usepackage[latin1]{inputenc} 
\usepackage{color}
\usepackage{lmodern}
\usepackage{balance}
\usepackage[plainpages=false,pdfpagelabels,colorlinks=true,linkcolor=blue,urlcolor=blue,citecolor=blue,pdftitle={Expansion dynamics of interacting bosons in homogeneous lattices in one and two dimensions},pdfauthor={},pdfdisplaydoctitle=true,pdfduplex=DuplexFlipLongEdge]{hyperref}

\begin{document}

\title{Expansion Dynamics of Interacting Bosons in Homogeneous Lattices\protect\\in One and Two Dimensions}

\author{J. P. Ronzheimer} \affiliation{Department of Physics, Ludwig-Maximilians-Universität München, 80799 München, Germany}\affiliation{Max-Planck-Institut für Quantenoptik, 85748 Garching, Germany}
\author{M. Schreiber} \affiliation{Department of Physics, Ludwig-Maximilians-Universität München, 80799 München, Germany}\affiliation{Max-Planck-Institut für Quantenoptik, 85748 Garching, Germany}
\author{S. Braun} \affiliation{Department of Physics, Ludwig-Maximilians-Universität München, 80799 München, Germany}\affiliation{Max-Planck-Institut für Quantenoptik, 85748 Garching, Germany}
\author{S. S. Hodgman} \affiliation{Department of Physics, Ludwig-Maximilians-Universität München, 80799 München, Germany}\affiliation{Max-Planck-Institut für Quantenoptik, 85748 Garching, Germany}
\author{S. Langer} \affiliation{Department of Physics and Arnold Sommerfeld Center for Theoretical Physics,
Ludwig-Maximilians-Universität München, 80333 München, Germany}\affiliation{Department of Physics and Astronomy, University of Pittsburgh, Pittsburgh, Pennsylvania 15213, USA}
\author{I. P. McCulloch} \affiliation{Centre for Engineered Quantum Systems, School of Mathematics and Physics, The University of Queensland, St. Lucia, QLD 4072, Australia}
\author{F. Heidrich-Meisner} \affiliation{Department of Physics and Arnold Sommerfeld Center for Theoretical Physics,
Ludwig-Maximilians-Universität München, 80333 München, Germany}\affiliation{Friedrich-Alexander Universit\"at Erlangen-N\"urnberg, Institut f\"ur Theoretische Physik II, 91058 Erlangen, Germany}
\author{I. Bloch} \affiliation{Department of Physics, Ludwig-Maximilians-Universität München, 80799 München, Germany}\affiliation{Max-Planck-Institut für Quantenoptik, 85748 Garching, Germany}
\author{U. Schneider} \affiliation {Department of Physics, Ludwig-Maximilians-Universität München, 80799 München, Germany}\affiliation{Max-Planck-Institut für Quantenoptik, 85748 Garching, Germany}

\vskip 0.25cm

\date{\today}

\begin{abstract}

We experimentally and numerically investigate the expansion of initially localized ultracold bosons in 
homogeneous one- and two-dimensional optical lattices. We find that both dimensionality and interaction 
strength crucially influence these non-equilibrium dynamics. While the atoms expand ballistically in 
all integrable limits, deviations from these limits dramatically suppress the 
expansion and lead to the appearance of almost bimodal cloud shapes, indicating diffusive dynamics in the center
surrounded by ballistic wings. 
For strongly interacting bosons, we observe a dimensional crossover of the dynamics from ballistic 
in the one-dimensional hard-core case to diffusive in two dimensions, as well as a similar 
crossover when higher occupancies are introduced into the system.

\end{abstract}

\pacs{}
\maketitle

Non-equilibrium dynamics of strongly correlated many-body systems pose one of
the most challenging problems for theoretical physics~\cite{Polkovnikov2011}. Especially in one dimension, 
many fundamental questions concerning transport properties and relaxation dynamics in isolated
systems remain under active debate. These problems have attracted a renewed interest 
in recent years due to the advent of ultracold atomic gases. The ability to control various system parameters in real time
has not only allowed quantum simulations of equilibrium properties
of interacting many-body systems~\cite{Bloch2008}, but has also enabled experimental studies of quantum
quenches~\cite{Greiner2002,Sadler2006,Kinoshita2006,Chen2011,Trotzky2012} and particle transport~\cite{Ott2004,Fertig2005,Strohmaier2007,Schneider2012,Brantut2012} in clean, well-controlled, and isolated systems.
Here, we study the combined effects of interactions and dimensionality on the expansion dynamics of bosonic atoms in optical lattices.

\begin{figure}
\includegraphics{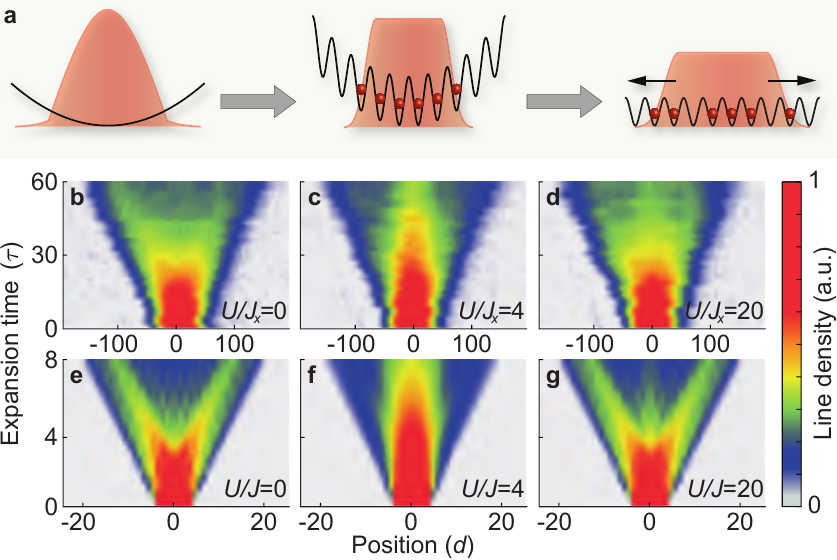}
\caption{\label{fig:Sketch}{Experimental sequence and time evolution during the expansion.} (a) Sketch of the experimental sequence. (b)-(d) Experimental time evolution of line density profiles during a 1D expansion for various interaction strengths (each line is individually normalized). (e)-(g) Corresponding t-DMRG calculations for eight atoms, plotted using cubic interpolation.}
\end{figure} 

While interactions generally lead to diffusive transport in higher dimensions, the situation is more involved in one dimension, where the phase space available for scattering can be severely limited. This was demonstrated, for example, by the experimental realization of a quantum Newton's cradle~\cite{Kinoshita2006}, showing that not all 1D Bose gases thermalize (see also \cite{Gring2012}). An intriguing phenomenon in one dimension is the existence of an exact mapping~\cite{Cazalilla2011} from hard-core bosons on a lattice or a Tonks-Girardeau gas~\cite{Paredes2004, Kinoshita2004c} to non-interacting spinless fermions, demonstrating the integrability of these systems. Furthermore, this mapping establishes that the time evolution of the density distribution is identical for hard-core bosons and non-interacting fermions. As a consequence, hard-core bosons in one dimension expand ballistically and, asymptotically, undergo a dynamical fermionization during the expansion~\cite{Rigol2005a,Minguzzi2005a}. 
In a transient regime, 
even initial 1D Mott insulators with unity filling are predicted to become coherent during the expansion
and to dynamically form long-lived quasi-condensates at finite momenta~\cite{Rigol2004b, Rodriguez2006a,[The dynamical emergence of coherence was also observed for fermions in: ]Heidrich-Meisner2008}.
In the presence of doubly occupied lattice sites (doublons) or even higher occupancies, the above mapping is not applicable. The dynamics then become more involved and can include intriguing quantum distillation effects, namely a demixing of doublons and single atoms~\cite{Heidrich-Meisner2009,Muth2012}.

Several powerful theoretical methods have been used to study the expansion dynamics in one dimension, including the
time-dependent density matrix renormalization group method (t-DMRG) (see, e.g., \cite{Rodriguez2006a,Heidrich-Meisner2009,Kajala2011}) and approaches based on the existence of exact solutions 
(see, e.g., \cite{Iyer2012,Caux2012,Ohberg2002,Campo2006,Jukic2009}). 
For interacting 2D systems, in contrast, one needs to resort to approximate methods such as the time-dependent Gutzwiller
ansatz, which predicts dynamical condensation even in two dimensions \cite{Jreissaty2011,Wernsdorfer2012}. %hen10,,Wernsdorfer2010

In this work, we experimentally study the expansion of 
initially localized bosonic atoms in the lowest band of an optical lattice.
We investigate how the expansion speed changes as a function of interaction strength and 
how it is affected by the dimensionality of the system. Furthermore, we identify the role
of multiply occupied lattice sites in the system and compare our results to t-DMRG \cite{Vidal2004,Daley2004,White2004} calculations in the 1D case.

\emph{ Experimental sequence --- } The experiment starts with a Bose-Einstein condensate of approximately $10^5$
bosonic $^{39}$K atoms in a three-beam optical dipole trap. The condensate  is loaded into a blue-detuned, three-dimensional optical lattice 
(lattice constant $d=\lambda/2$, wavelength $\lambda=736.7\,\mathrm{nm}$) 
with a lattice depth of $V_0=33.0\!\left(5\right)\, E_{{r}}$. Here, $E_{{r}}=h^2/\left(2m\lambda^2\right)$ 
denotes the recoil energy, $m$ the atomic mass, and $h$ is 
Planck's constant. For suitable harmonic confinements, sufficiently strong repulsive interactions, and adiabatic loading, a large Mott insulating core with unity filling and a radius of $\left(40-50\right)d$ is created in the center (see Fig.~\ref{fig:Sketch}(a)).
By employing a \hyphenation{Fesh-bach}Feshbach resonance at a magnetic field of $402.50\!\left(3\right)\,\mathrm{G}$  
we can tune the interaction strength during loading and thereby control the amount of multiply occupied lattice sites. In the deep lattice, where tunneling is suppressed (tunneling time $\tau_{{d}} = \hbar / J_{{d}} \approx 58\, \mathrm{ms}$, with the tunneling amplitude $J_{d}$ and $\hbar=h/(2\pi)$), the atoms are held for a $20 \, \mathrm {ms}$ dephasing period, during which any residual coherences between lattice sites are lost \cite{Will2010} and all atoms become localized to individual lattice sites. The resulting state after this loading procedure is a product of local Fock states, $\left|\Psi_{\mathrm{initial}}\right\rangle=\prod_i{ \frac{1}{\sqrt{\eta_i!}}\left( \hat{b}_i^{\dagger}\right)^{\eta_i}}\left|0\right\rangle,\; \eta_i\in\{0,1,2,\dots\}$, where $\hat{b}_i^{\dagger}$ is the creation operator for a boson on site $i$. This state is characterized by a flat quasimomentum distribution $n_k=\mathrm{const}$, where $k\in[-\pi/d,\pi/d]$ denotes the quasimomentum. During the dephasing period, we change the magnetic field to set the desired interaction strength $U/J$ for the expansion. Due to the suppressed hopping during this part of the sequence, this field ramp does not alter the density distribution, i.e., the initial state prior to the expansion is identical for all interactions.
The expansion is initiated by lowering the lattice depth along one or both horizontal directions ($x,y$) in $150\,\mu\mathrm{s}$ to a depth of $8.0\left(1\right)\,E_{r}$ to induce tunneling with amplitudes $J_x$ ($\tau = \hbar / J_{x} = 0.55\,\mathrm{ms}$) and $J_y$ between neighboring lattice sites along these directions. This is equivalent to a quantum quench from $U/J \approx \infty$ to a finite $U/J$. Simultaneously, the strength of the dipole trap is reduced to a small but finite value that compensates the anti-confinement along the expansion direction created by the lattice beams (see \cite{SOM} for supplementary information). 

The dynamics in the resulting lattice can be described within the homogeneous Bose-Hubbard model: 
\begin{equation*}
H=-J_x \sum_{\langle i,j \rangle_x}{\hat{b}_i^{\dag}{\hat{b}_j}}-J_y \sum_{\langle i,j \rangle_y}{\hat{b}_i^{\dag}{\hat{b}_j}}+\frac{U}{2}\sum_{i}{\hat{n}_i\left(\hat{n}_i-1\right).}
\end{equation*}
Here, $U$ denotes the on-site interaction strength, $\hat{n}_i=\hat{b}_i^\dagger \hat{b}_i$, and $\langle i,j \rangle_{x(y)}$ indicates a summation over nearest neighbors along the $x$- ($y$-)direction. 
\begin{figure}
\includegraphics{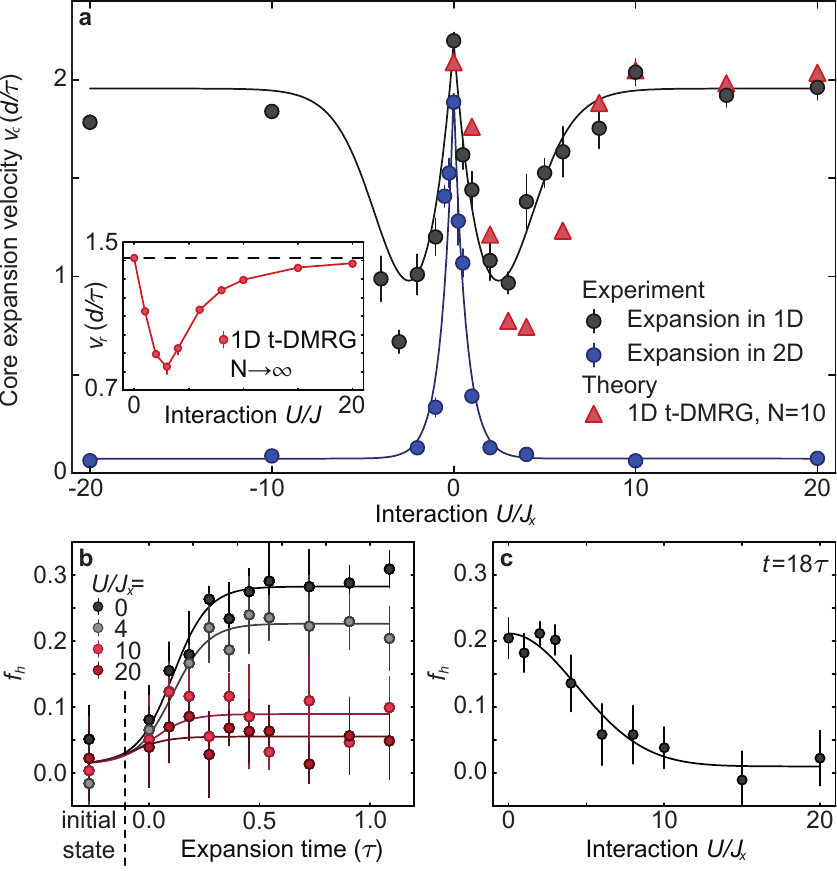}
\caption{\label{fig:1D_Expansion}{Core expansion velocity and dynamical generation of higher occupancies.} (a) Core expansion velocity $v_{c}$ for experimental data in one dimension (black circles, lattice depth $\left(8,33,33\right)E_{r}$ along $\left(x,y,z\right)$, $J_y\approx 0$) and two dimensions (blue circles, $\left(8,8,33\right)E_{r}$, $J_x=J_y$) and t-DMRG calculations for $N=10$ particles in one dimension (red triangles). Experimental error bars denote the standard deviation of the linear fits. Inset: $v_{r}$ calculated by t-DMRG and extrapolated to infinite particle number. Error bars are given by the uncertainty of the extrapolation~\cite{SOM}. (b) Higher occupancy, as measured by $f_{h}$,  versus expansion time in the experiment. For the points labeled "initial state", the measurement was performed directly after the dephasing period in the deep lattice~\cite{SOM}. (c) $f_{h}$ after an expansion time of 
$t=18\,\tau$. Error bars in (b) and (c) show standard deviations of averaging four data points. All lines are guides to the eye.}
\end{figure}

We monitor the \textit{in situ} density distribution of the expanding cloud using standard absorption imaging along the vertical axis.
The recorded column densities are integrated over one direction and the resulting line densities are presented in Figs.~\ref{fig:Sketch}(b-d) as a function of the expansion time for the 1D case. 
In both the non-interacting and the hard-core limits we expect a ballistic expansion
which splits the cloud into a left- and right-moving portion~\cite{Rodriguez2006a,Langer2012,Polini2007}, as can be seen in our numerical results shown in Figs.~\ref{fig:Sketch}(e,g). 
While the splitting can be clearly observed in the experimental data for the non-interacting case (Fig.~\ref{fig:Sketch}(b)), the presence of a few multiply occupied lattice sites decreases its visibility in the strongly interacting case (Fig.~\ref{fig:Sketch}(d)).

\emph{ Expansion velocities ---} 
To quantify the expansion dynamics we extract the half-width-at-half-maximum (HWHM) 
from the line density profiles~\footnote{In the case of a double peak structure (see, e.g., Fig 1(e)), the HWHM measures half of the distance between the outer edges of the two peaks~\cite{SOM}} and determine the core expansion velocities $v_{c}$ (Fig. 2(a)) via linear fits to the
evolution of the HWHM at intermediate times~\cite{SOM}.
In both one and two dimensions, the maximum core expansion velocity occurs in the non-interacting limit, 
where the system expands ballistically. Due to an exact dynamical symmetry of Hubbard models on bi-partite lattices, the expansion dynamics are independent of the sign of the interaction \footnote{The proof for this dynamical symmetry given in~\cite{Schneider2012} carries over to the bosonic case} and we therefore focus the discussion on the $U>0$ case.
In two dimensions, increasing the interaction strength monotonically reduces the core expansion velocity until it essentially drops to zero.
In one dimension, in contrast, a similar but much weaker suppression of the expansion velocity extends only up to interaction strengths on the order of the bandwidth $U\sim 4J_x$, while $v_{c}$ increases again for stronger interactions and eventually reaches values comparable to the non-interacting case. 

The same qualitative behavior is evident in the t-DMRG simulations for ten particles, shown as red triangles in Fig.~\ref{fig:1D_Expansion}(a). Since the numerically calculated HWHM suffers from rather large finite-size effects, 
we also present t-DMRG results for an alternative measure of the expansion velocity, namely $v_{r}=(d/dt) \sqrt{R^2{\left(t\right)}-R^2{\left(0\right)}}$, extracted from the radius 
$R^2(t)=(1/N)\sum_{i} \langle \hat n_i(t)\rangle (i-i_0)^2 d^2$ (inset), where $N$ is the particle number and $i_0$ denotes the central lattice site. It is more robust against finite-size effects and allows an extrapolation to infinite particle number~\cite{SOM}, and, in our setup, exhibits the same qualitative dependence on $U$. Moreover, at $U=0$, $v_{r}$ has an intuitive physical interpretation, as it is in this case equal to the average expansion velocity $v_{\mathrm{av}}$. The latter is given by the initial quasimomentum distribution through $v_{\mathrm{av}} = 1/(N\hbar)\sqrt{\sum_k (\partial\epsilon_k/\partial k)^2 n_k}$, where $\epsilon_k =-2J\cos(kd)$ denotes the tight-binding dispersion relation. For the given initial state, where $n_k$ is flat, this results in $v_{\mathrm{av}}=\sqrt{2}\left(d/\tau\right)$, illustrated by the dashed line in the inset of Fig.~\ref{fig:1D_Expansion}(a). 
Usually, one would associate a constant velocity with a ballistic expansion and would expect $\sqrt{R^2{\left(t\right)}-R^2{\left(0\right)}} \propto \sqrt{t}$ for diffusive dynamics. In the case of the sudden expansion, however, the interpretation is more complicated, because the diffusion constant is density dependent and the density distribution is inhomogeneous and time-dependent (see~\cite{Schneider2012} and~\cite{Langer2012} for details).

The fast expansion for strong interactions in one dimension is a consequence of the system entering into the hard-core boson regime, where, at $U=\infty$, it can be exactly mapped to non-interacting fermions, which expand ballistically with $v_{r}=\sqrt{2}\left(d/\tau\right)$~\cite{SOM}. 
Even though hard-core bosons undergo collisions and their quasimomentum distribution changes over time~\cite{Rigol2004b, Rodriguez2006a}, the above mapping guarantees that the evolution of their density distribution is ballistic and identical to the non-interacting case. 
In other words, the conservation of the quasimomentum distribution of the underlying non-interacting fermions severely constrains the scattering processes, thereby preventing the dynamics from becoming diffusive.

Starting from the hard-core boson limit, the decrease of the expansion velocity towards smaller interactions can be qualitatively understood by considering the dynamical formation of doublons and higher occupancies. For $U/J \gtrsim 4$, isolated doublons in one dimension can be thought of as heavy compound objects, propagating with typical effective hopping matrix elements on the order of $J^2/U$ \cite{Folling2007}. While their formation is energetically suppressed at $U/J \gg 4$, for smaller $U$ the system can maximize its local entropy through the formation of doublons (and higher occupancies) during the early phase of the expansion (see Figs.~\ref{fig:1D_Expansion}(b,c)). Therefore, as $U$ decreases, higher occupancies begin to form and the expansion velocity decreases. In addition, the possibility of creating higher occupancies increases the phase space available for scattering and therefore favors diffusive dynamics. For vanishing interactions, the scattering cross section approaches zero and the expansion becomes ballistic again with a large velocity of $v_{r} = \sqrt{2} (d/\tau)$. Therefore, there has to be a minimum of $v_{c}$ at some intermediate $U$, which turns out to be close to the critical $U/J \approx 3.4$ for the 1D Mott insulator to
superfluid transition~\cite{Kuhner2000}. This is consistent with other studies of
quantum quenches, which observe the fastest relaxation
times close to the critical point~\cite{Trotzky2012,Cramer2008}. 

The buildup of higher occupancies during the initial expansion dynamics shown in Fig.~\ref{fig:1D_Expansion}(b) is monitored by comparing the number of atoms left after a parity projection $N_{\mathrm{par}}$ with the total atom number $N_{\mathrm{total}}$, yielding $f_{h}=\left(N_{\mathrm{total}}-N_{\mathrm{par}}\right)/N_{\mathrm{total}}$.
In the absence of triply or higher occupied sites, $f_{h}$ measures the fraction of atoms on doubly occupied sites~\cite{SOM}.
While the expansion starts from an initial state with essentially no higher occupancy, $f_{h}$ rises significantly over roughly the first half tunneling time. After this initial buildup, $f_{h}$ remains almost constant and changes only on the much slower timescale of the expansion (compare Fig.~\ref{fig:1D_Expansion}(c)).
This initial fast relaxation is purely local, as can be seen in t-DMRG calculations comparing the relaxation timescale to the evolution of the system without opening the trap~\cite{SOM}. The formation of higher occupancies is accompanied by changes in $n_k$ and results in an increase of interaction energy and therefore a decrease in kinetic energy. The effect of the reduced kinetic energy (as measured by $v_{\mathrm{av}}$) is, however, much smaller than the observed reduction of the expansion velocity~\cite{SOM}. We thus conclude that
scattering processes during the expansion are mainly responsible
for the slower expansion.

\begin{figure}
\includegraphics{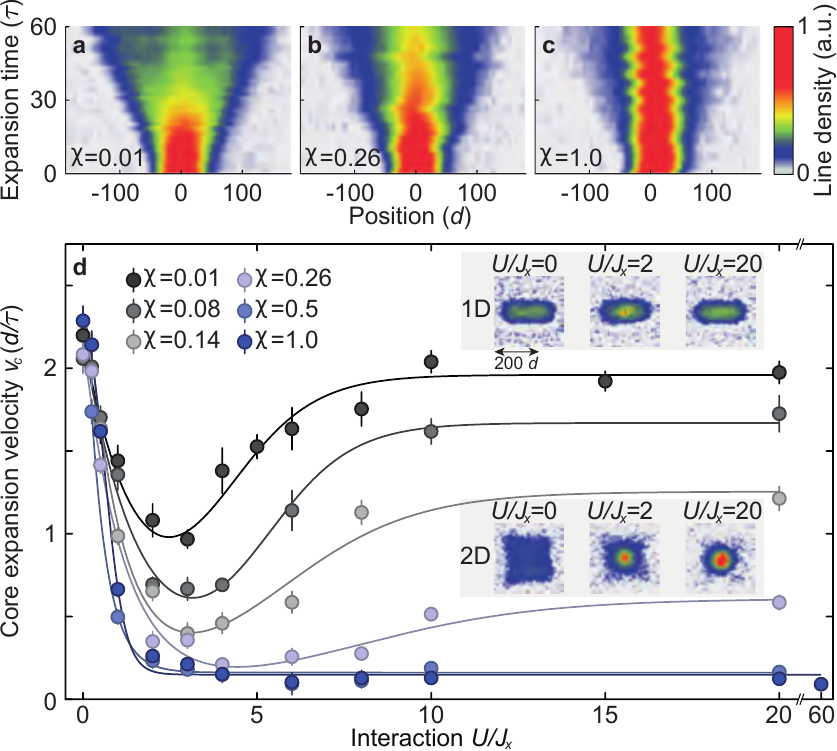}
\caption{\label{fig:1D_2D_crossover}{1D-2D Crossover.} (a)-(c) Evolution of line density profiles for various tunneling ratios $\chi=J_y/J_x$ and $U/J_x=10$. (d) Experimental core expansion velocity $v_{c}$ for various $\chi$. Lines are guides to the eye. Error bars denote the standard deviation of the linear fits. The insets show the column density at $t\approx 36\,\tau$.}
\end{figure}

\emph{ 1D-2D crossover --- } 
In Fig.~\ref{fig:1D_2D_crossover} we analyze how the expansion dynamics change when we gradually tune the dimensionality from a purely 1D system towards a 2D geometry. This is implemented by varying the depth of the lattice along the $y$-direction and thereby the tunneling ratio $\chi= J_y/J_x$ for the expansion \footnote{The strength of the dipole trap was adjusted as well in order to guarantee a flat potential along the $x$-axis}. Upon increasing $\chi$, the expansion dynamics at strong interactions change fundamentally. Instead of the fast expansion observed in the 1D case (Fig.~\ref{fig:1D_2D_crossover}(a)), the major fraction of the cloud simply remains in the center (Fig.~\ref{fig:1D_2D_crossover}(c)). Moreover, the column density profiles shown in the insets of Fig.~\ref{fig:1D_2D_crossover}(d)
exhibit a characteristic bimodal structure.
In the 2D case, this structure consists of a slowly expanding, round, diffusive core on top of a
square-shaped ballistic background and can be seen for all moderate to strong interactions.
In one dimension, on the other hand, a similar behavior is only visible for intermediate interaction strengths.

In Fig.~\ref{fig:1D_2D_crossover}(d), we illustrate how the interaction dependence of $v_{c}$  changes as we go from a 1D system with two integrable limits to a 2D system, where only the non-interacting case is integrable. The expansion speed in the non-interacting case is independent of $\chi$, since in this case the dynamics along the two lattice axes are separable. For all values of $\chi$, the expansion speed initially decreases with increasing interactions.  
For small $\chi$, the core expansion velocity increases again for strong interactions, whereas, for $\chi>0.5$, it remains minimal. The behavior at large $\chi$, as well as the bimodal cloud shape, is analogous to the dynamics of strongly interacting lattice fermions in two dimensions, which were shown to be diffusive~\cite{Schneider2012}. 
The square-shaped background consists of ballistically expanding atoms originating from the edge of the high density core, while collisions render the expansion diffusive for atoms inside the core. Such diffusive dynamics are consistent with the numerical observation that hard-core bosons in two dimensions thermalize~\cite{Rigol2008}.
Our experimental results show a qualitative difference between the dynamics in one and two dimensions in the strongly interacting regime, whereas theoretical studies using the time-dependent Gutzwiller ansatz predict a qualitatively similar behavior, independent of dimension~\cite{Jreissaty2011,Itay2010,Wernsdorfer2012}. Overall, we observe that, for interacting systems, the expansion along one direction is suppressed by an increased tunneling along a transverse direction. This promotes the notion that increasing transverse tunneling enlarges the accessible phase-space for scattering processes and therefore favors diffusive dynamics.

\begin{figure}
\includegraphics{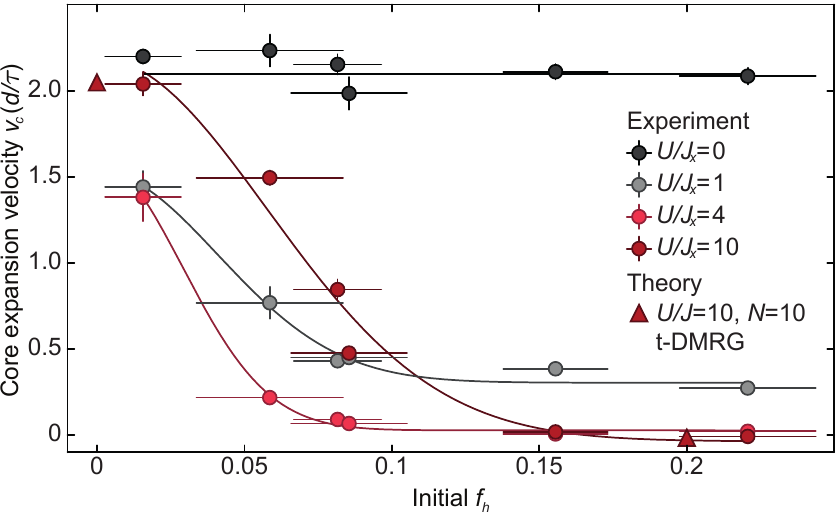}
\caption{\label{fig:Expansion_vs_initial_DO}{Effect of higher occupancies in the initial state.} Core expansion velocity $v_{c}$ in one dimension as a function of the higher occupancy, as measured by $f_{h}$, in the initial state for various interaction strengths. Solid lines are guides to the eye. Vertical error bars show the standard deviation of the linear fits, horizontal error bars the standard deviation of averaging $16$ measurements.}
\end{figure}

\emph{ Higher occupancies in the initial state ---} Figure~\ref{fig:Expansion_vs_initial_DO} illustrates the effect of a random admixture of higher occupancies in the initial state on the expansion dynamics. 
This admixture is created by loading the lattice at smaller interaction strength and higher densities, such that no clear Mott insulator will form. Nonetheless, the dephasing in the deep lattice remains effective, such that the initial state of the expansion can still be described as a product of local Fock states, but with higher occupancies on some randomly chosen sites. 
While there is, as expected, no significant effect of multiply occupied sites
in the non-interacting case, where each atom expands individually, already at $U/J_x = 1$ higher occupancies in the initial state reduce the core expansion velocity. This reduction becomes most dramatic close to the hard-core limit ($U/J_x = 10$), where the originally high expansion velocity quickly approaches zero.
In this limit, any higher occupancies are long-lived~\cite{Winkler2006} and their small effective higher-order tunneling rate slows down the expansion~\cite{Heidrich-Meisner2009,Kajala2011}. Furthermore, the presence of multiply occupied lattice sites in the strongly interacting limit can give rise to quantum distillation processes~\cite{Heidrich-Meisner2009} and thereby the formation of a stable core of doubly occupied lattice sites \cite{Petrosyan2007}.

\emph{ Conclusion --- }
Experimentally, we find the fastest expansions near the exactly solvable limits of the Bose-Hubbard model, where additional conservation laws restrict scattering such that diffusion is not possible. These are: (i) the non-interacting limit, irrespective of dimension, and (ii) 
the case of infinitely strong interactions in one dimension, provided there are no higher occupancies in the initial 
state. Deviations from these cases, either by finite interactions, the crossover towards two dimensions, or an admixture of 
higher occupancies in the initial state, lead to a substantial suppression of the expansion. 
In the case of the crossover to two dimensions at large $U/J$, the emergence of diffusive dynamics in the core is additionally signaled by the characteristic bimodal cloud shape previously observed in the fermionic case~\cite{Schneider2012}. 
In one dimension at intermediate interactions or with initially multiply occupied lattice sites, both experimental 
and t-DMRG profiles suggest an almost bimodal structure here as well. 
Therefore, we conjecture that the common reason for the slow expansions seen in the experiments is the 
emergence of diffusive dynamics in the core region of the cloud. 

\vspace{5 mm}

We thank Daniel Garbe and Tim Rom for their contributions in constructing the experimental apparatus and Lode Pollet, Marcos Rigol, Achim Rosch, and Julia Wernsdorfer for insightful discussions. We acknowledge financial support from the Deutsche
Forschungsgemeinschaft (FOR801, FOR912, Deutsch-Israelisches
Kooperationsprojekt Quantum phases of ultracold atoms
in optical lattices), the US Defense Advanced Research
Projects Agency (Optical Lattice Emulator program),
and Nanosystems Initiative Munich.

\bibliographystyle{apsrev4-1}
\bibliography{2013_01_21_Expansion}
\vspace{2cm}
{\Large \centering \bf Supplemental Material \\}

\renewcommand{\thefigure}{S\arabic{figure}}
\renewcommand{\theequation}{S\arabic{equation}}

\section{Experimental details}

\subsection{Preparation of the initial state}

We prepare a condensate of approximately $10^5$
%$N= 70\left(5\right) \times 10^{3}$ 
$^{\mathrm{39}}$K atoms in the $\left|F=1,m_F=1\right>$ hyperfine state in a three-beam optical dipole trap with trap frequencies of
$\omega_{x}= \omega_{y}= 2\pi \times 52\!\left(2\right)\,\mathrm{Hz}$ along the horizontal ($x,y$) directions and $\omega_{z} =2\pi \times 119\!\left(8\right) \,\mathrm{Hz}$ along the
vertical ($z$) direction. For the experiments using an initial state of almost exclusively singly occupied sites, presented in Figs.~2 and 3 of the main text, the initial scattering length is set to $a_{{s}}= 350\!\left(7\right)\,a_0$, where $a_0$ is the Bohr radius, by employing a Feshbach resonance at $402.50\!\left(3\right)\,\mathrm{G}$ \cite{Zaccanti2009}.
We linearly ramp up the lattice potential (lattice constant $d=\lambda/2=368.3\,\mathrm{nm}$) in $8\,\mathrm{ms}$ to a depth of $20.0\!\left(3\right)\,E_{{r}}$. We freeze out the resulting density distribution by a second lattice ramp to
$33.0\!\left(5\right)\, E_{{r}}$ in $1\,\mathrm{ms}$. At the same time we turn off the vertical confinement by switching off the dipole trap beams
along the horizontal axes. The atoms are now suspended against gravity solely by the deep vertical lattice, which remains at this depth during the rest of the experiment.
The small tunneling amplitude of $J_d=h \times 2.7\!\left(2\right) \,\mathrm{Hz}$ in the deep lattice, in combination with the effects of gravity, induces Bloch oscillations with an oscillation length of $11\!\left(1\right)\,\mathrm{nm}$ that is small compared to the lattice constant. The system is thereby effectively decoupled into independent 2D systems.
The intensity of the vertical dipole trap beam is increased simultaneously with the loading of the lattice to compensate the increasing anti-confinement caused by the lattice beams. In the deep lattice, the combined trap frequencies along the horizontal 
directions are $\omega_{x}= 2\pi \times 56\!\left(6\right)\, \mathrm{Hz}$ and $\omega_{y}= 2\pi \times 50\!\left(5\right)\,\mathrm{Hz}$.
The atoms are held in the deep 3D lattice for $20\,\mathrm{ms}$ while the magnetic field is ramped to set the scattering length for the expansion.
\begin{figure}
\includegraphics{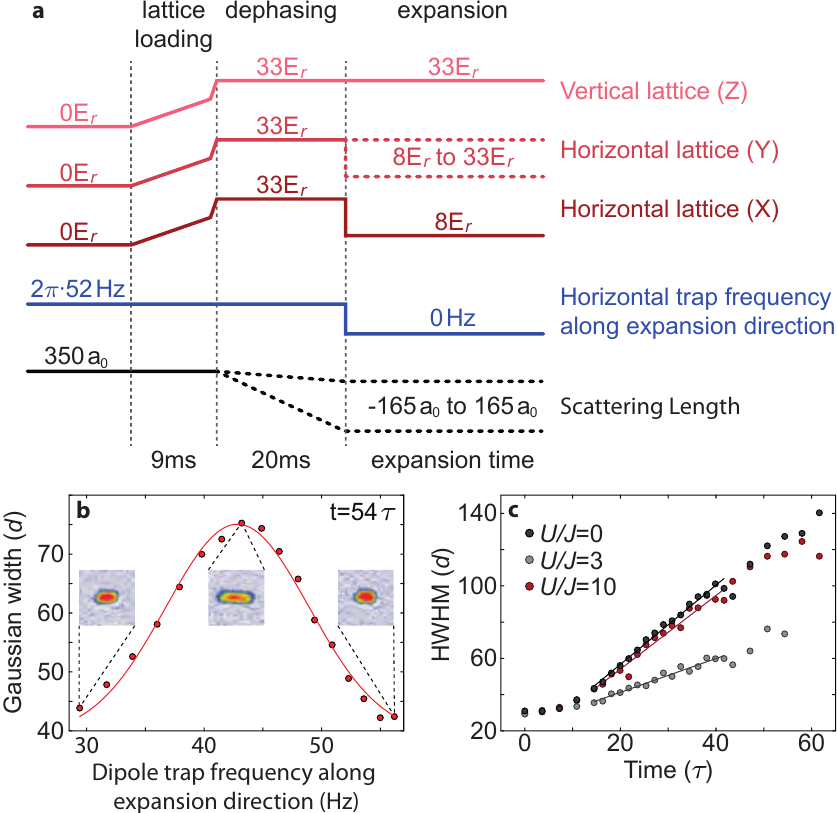}
\caption{\label{fig:Dipole_and_lin_fits}{Experimental sequence, optimization of homogeneity and extraction of $v_{{c}}$.} (a) Sketch of the experimental sequence (not to scale). (b) Width of Gaussian fit to line density profiles after an expansion time of $t=54\,\tau$ at $U/J_x \approx 56$ as a function of the trap frequency due to the vertical dipole trap beam, neglecting the anti-confinement due to the lattice beams. Insets show the column 
density distribution at the indicated points. 
(c) Extracted HWHM of expanding clouds in one dimension for various interaction strengths. The lines show the fit result for $v_{{c}}$ and their extension indicates the fit range.}
\end{figure}  
During this period, all coherences between lattice sites are lost and all atoms become localized to individual lattice sites \cite{Will2010}. The small tunneling rate ensures that atoms cannot redistribute during this hold time such that the resulting state is identical for all final interaction strengths.
The expansion is initiated by ramping down the lattice depth along one or two horizontal directions and simultaneously adjusting the intensity of the remaining
vertical dipole trap beam within $150\,\mu\mathrm{s}$, a timescale that is slow enough to avoid excitations 
into higher bands but fast compared to the tunneling rate. Figure~\ref{fig:Dipole_and_lin_fits}(a) presents a sketch of the intensity and field ramps employed in the experimental sequence. For the data with an increased density of higher occupancies (Fig.~4), we reduce the initial interaction strength and increase the initial trap frequencies, such that no clear Mott insulator can form.

\subsection{Optimization of homogeneity}

The optical potentials during the expansion are provided by the three blue-detuned optical lattices, with Gaussian waists of approximately $150\,\mu\mathrm{m}$, and the red-detuned dipole trap beam along the vertical
axis with approximately the same Gaussian waist. In the overlap region of these beams, the anti-confining potential due to the lattice beams can thus be compensated along the horizontal directions
by the confining potential of the dipole trap beam. An exact compensation along both horizontal directions is possible for equal lattice depths along these directions. In all other cases, we optimize the
compensation for the $x$-direction, along which we record the dynamics. To perform the optimization, we let the atoms expand in the lattice for a fixed expansion time and various intensities of the dipole trap beam and maximize the final size of the cloud (see Fig.~\ref{fig:Dipole_and_lin_fits}(b)). Note that both confining as well as anti-confining potentials hinder the expansion~\cite{Schneider2012}.

\subsection{Determination of the core expansion velocity $v_{{c}}$}\label{Exp_det_vc}

To determine the core expansion velocity $v_{{c}}$ along the $x$-direction, we first integrate the recorded \textit{in situ} column densities of the clouds along the transverse ($y$) direction to obtain line density profiles. For each of these profiles, we determine 
the maximum density $n_{\mathrm{max}}$. Starting from the outer edges of the profiles, we move inwards in both directions and determine the positions where the density first reaches $n_{\mathrm{max}}/2$, using linear interpolation between the points. Half of the distance between these two positions is recorded as the half-width-at-half-maximum (HWHM) of the
cloud. In the case of a double peak structure (see, e.g., Fig. 1(e) in the main text) the HWHM measures half of the distance between the outer edges of the two peaks. Typical time evolutions of the HWHM are shown in Fig.~\ref{fig:Dipole_and_lin_fits}(c). Even in the non-interacting cases, the HWHM does not significantly increase during the first  
few tunneling times of the evolution. 
The HWHM only grows significantly once the extension of the expanding single
particle wavefunctions becomes comparable to the initial cloud size~\cite{Campo2006}. For very large clouds, the assumption of a homogeneous lattice with
constant lattice depth is not valid anymore, because lattice and dipole trap beams have only a finite width, giving rise to residual potentials. Thus, we apply a linear fit with offset to the HWHM evolution only in the time range from $t\approx 14\,\tau$ to $t\approx42\,\tau$. The slope of this fit is the core expansion velocity $v_{{c}}$.

\subsection{Error estimates for $U$ and $J_x$}
Error bars for $U/J_x$ and $\tau$ are not given in Figs. 2 (a-c) and 3 (d) of the main text, as their statistical errors, caused by uncorrelated fluctuations, are much smaller than the width of the points. There are, however, sources of systematic errors
 for $U$ and $J_x$. From measurements of the expansion speed of non-interacting bosons over several days, we estimate the long term fluctuations of our calibration of $J_x$ to be on the order of $3\%$. 
The main contribution to the uncertainty in $U$ originates from the uncertainty of the width of the Feshbach resonance that is used to calculate the scattering length $a_{{s}}$ at a given magnetic field. The resulting uncertainty of the scattering length ranges from $0.5\,a_0$ at a set value of $0\,a_0$ to $3.8\,a_0$ at $188\,a_0$.

\subsection{Detection of multiply occupied sites}

\begin{figure}
\includegraphics{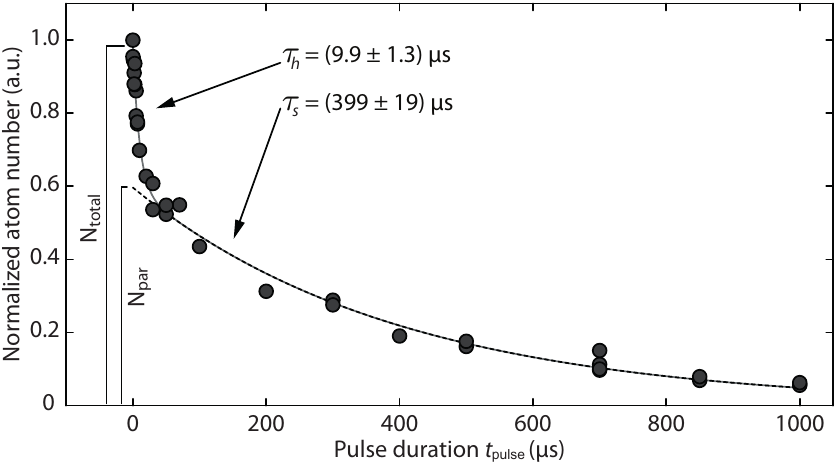}
\caption{\label{fig:Imaging_DO}{Detection of multiply occupied sites.} Loss of atoms for off-resonant light pulses of varying duration. The fast loss ($\tau_{{h}}$) originates from higher occupancies, the slow loss ($\tau_{{s}}$) is caused by off-resonant excitations.}
\end{figure}

In order to determine the fraction of atoms on multiply occupied sites, we first freeze out the on-site number distribution by ramping up the lattice in $50\,\mu\mathrm{s}$ to a depth of $33.0\!\left(5\right)\, E_{{r}}$ along all three axes. 
In the deep lattice, where tunneling is strongly suppressed (tunneling time  $\tau_{{d}} = \hbar / J_{{d}} \approx 58\, \mathrm{ms}$), we set the magnetic field within $10\,\mathrm{ms}$ to a fixed value of $B\approx 400\,\mathrm{G}$, where the scattering length is large. 
We then apply a near-resonant light pulse which is approximately $110\,\mathrm{MHz}$ red detuned relative to the high-field imaging transition from the $\left|4^2 S_{1/2}, m_I=+3/2, m_J=-1/2\right\rangle$ state to the $\left|4^2 P_{3/2}, m_I=+3/2, m_J=-3/2\right\rangle$ state. 
On multiply occupied sites, this near-resonant light pulse gives rise to a fast two-body loss process caused by light-assisted inelastic scattering of atoms \cite{DePue1999}. 
This loss acts as a parity projection of the on-site atom number and results in the loss of all atoms for even atom numbers and the loss of all but one atom for odd atom numbers. 
In Fig.~\ref{fig:Imaging_DO}, we present a typical decay curve of the total atom number in the presence of the near-resonant light for varying pulse durations. 
The initial fast loss (lifetime $\tau_{{h}}=9.9\!\left(1.3\right)\,\mu\mathrm{s}$) stems from losses on multiply occupied sites. 
After the initial decay, we observe a much slower decay ($\tau_{{s}}=399\!\left(13\right)\,\mu\mathrm{s}$) that is caused by off-resonant excitations of the remaining single atoms. 
We extract a measure of the higher occupancy by comparing the number of atoms with ($N_{\mathrm{pulse}}$) and without ($N_{\mathrm{total}}$) a near-resonant light pulse with a duration of $t_{\mathrm{pulse}}=50\,\mu\mathrm{s}$. 
In the presence of the near-resonant pulses, the parity projection on multiply occupied sites has taken place and only atoms on singly occupied sites as well as the remaining atoms from sites with $\eta_i=3,5,\ldots$ are left in the system. 
The measured atom number $N_{\mathrm{pulse}}$ is then extrapolated to a pulse duration of $0\,\mu\mathrm{s}$ using the measured slow decay time $N_{\mathrm{par}}=N_{\mathrm{pulse}}\exp{\left(t_{\mathrm{pulse}}/\tau_{{s}}\right)}$. 
We then calculate an approximate measure of the fraction of atoms on multiply occupied sites: $f_{{h}}=\left(N_{\mathrm{total}}-N_{\mathrm{par}}\right)/N_{\mathrm{total}}$. 
Note that, strictly speaking, $f_{{h}}$ is only a lower bound on the fraction of atoms on multiply occupied lattice sites, because $N_{\mathrm{par}}$ also contains one atom per site with $\eta_i=3,5,\ldots$\:. 
While being exact for singly and doubly occupied sites, the measured fractions will be systematically too low whenever a significant amount of sites with $\eta_i \geq 3$ is present in the system. However, significant amounts of sites with an occupation of $\eta_i\geq3$ contribute only in the weakly interacting regime (see the discussion of numerical results in Sec.~\ref{sec:tdmrg_doublons}).

\section{Time dependent DMRG simulations for 1D systems}
We employ the adaptive time dependent density matrix renormalization group (t-DMRG) to carry out the time evolution \cite{Vidal2004,Daley2004,White2004,Schollwock2011},
using a Krylov-space based method to propagate the wavefunction in time \cite{Park1986,Hochbruck1997}.
To ensure convergence of the time dependent wavefunctions we enforce a threshold on the discarded weight \cite{Schollwock2005} per time-step $\delta t=1/(16J)$ of $10^{-4}$. 
Typically, this corresponds to using about $2000$ states at the longest expansion times.
Furthermore, we introduce a cutoff $N_{{b}}=3$ in the number of bosons per site, unless stated otherwise. 
For the initial state considered in our numerical simulations, where the initial density is limited to $\langle \hat n_i \rangle=2$ ($\hat n_i=\hat b_i^\dag \hat b_i$), increasing the cutoff to $N_{{b}}>3$ 
results in changes of only about $1\%$ for most of the quantities considered here. For $U/J\leq 2$, however, the quasimomentum distributions are calculated with $N_{{b}}=N$, since fluctuations in the local 
particle number are larger in this regime. 
In the two integrable limits, $U=0$ and $U=\infty$, we also use exact diagonalization to compute time dependent quantities.

The timescales that can be accessed with t-DMRG are limited due to the growth of entanglement that is encoded in the
time-evolved wavefunction \cite{Schollwock2011}. This growth depends on the type of non-equilibrium problem as well as on other factors such 
as particle number and system size. We want to reach timescales at which all particles participate in
the expansion, which requires the total system size to be substantially larger (roughly by a factor of four in our simulations) than the extension of the region with a finite density in the
initial state. Therefore, the particle number is restricted to $N\leq 14$.

\subsection{Initial state and measures for the expansion velocity}

 \begin{figure*}[t]
\includegraphics[width=\textwidth]{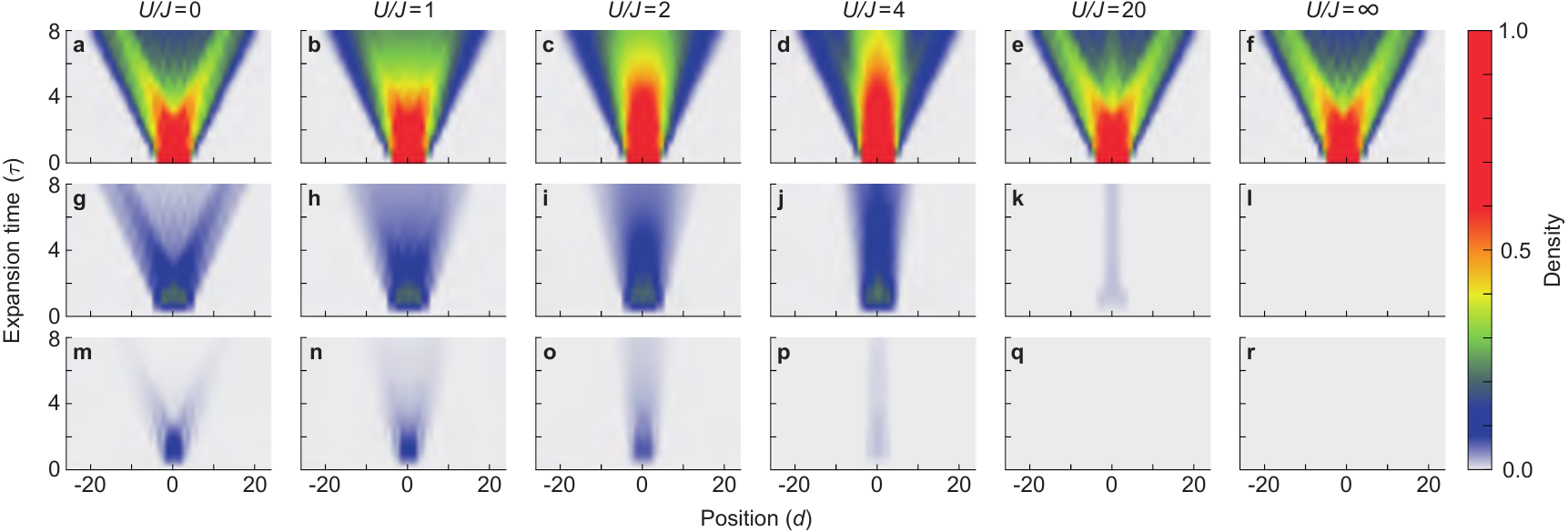}
\caption{{Time dependence of the density profile and of the probabilities for double and triple occupancy.} (a)-(f) 
Density profile $\langle {\hat n}_i(t)\rangle$, (g)-(l) double occupancy $\langle {\hat n}_{2,i}(t)\rangle$, (m)-(r) triple occupancy $\langle {\hat n}_{3,i}(t)\rangle$, for  the sudden expansion starting from a 
product of local Fock states with exactly one boson per site  ($N=8$, $U/J=0,1,2,4,20,\infty$). 
Panels (a), (g) and (m) show the results for free bosons, where the particles can form higher occupancies without energy cost. While the outer wavefronts in the time dependent particle density $\langle {\hat n}_i(t)\rangle$ are given by the maximum single-particle velocity of $2(d/\tau)$ for all $U$,  in panels (c) and (d) we observe 
a long-lived high density core significantly slowing down 
the expansion.   
The dynamics at $U/J=20$ (shown in (e), (k) and (q)) are very similar to the limit of hard-core bosons at $U/J=\infty$ ((f), (l) and (r)). }
\label{fig:epaps1}
\end{figure*}

The initial state for our simulations is a state
with exactly $\eta_i=0,1$ or $2$ bosons per site:
 \begin{equation}
|\psi_{\mathrm{initial}}\rangle=\prod_{i}\frac{1}{\sqrt{\eta_i!}}(\hat b^\dag_i)^{\eta_i}|0\rangle\,.
\label{eq:init}
\end{equation}
For the t-DMRG data shown in Figs.~1 and 2 of the main text and Fig.~\ref{fig:epaps1}, we consider a state with $\eta_i=1$ on $L$ adjacent sites and $\eta_i=0$ otherwise 
such that, in this case, the total particle 
number $N=\sum_i \langle \hat{n}_i\rangle $ equals $L$.
For this particular initial state, the kinetic energy
\begin{equation}
E_{\mathrm{kin}}=-J\sum_{\langle i,j\rangle} \langle \hat b^\dag_i \hat b_{j}\rangle
\end{equation} and the interaction energy
\begin{equation}
E_{\mathrm{int}}= \frac{U}{2}\sum_i \langle \hat n_i (\hat n_i-1)\rangle
\end{equation}
 both vanish and the initial quasimomentum distribution
 \begin{equation}
 n_k=\frac{1}{N}\sum_{l,m}e^{-ik(l-m)} \langle \hat b^\dag_l \hat b_m \rangle
 \end{equation}
  is flat, $n_k=\mbox{const}$.
The total energy $E=E_{\mathrm{kin}}+E_{\mathrm{int}}$, often referred to as the release energy,
is an integral of motion and its value is independent of $U$ in our problem.

In the experiment, a harmonic trapping potential exists during the lattice loading, and hence also regions of lower average density,
yet the particle numbers that we can run reliable simulations for are too small to account for this.

We study several measures of the expansion velocity: 
First,  the core expansion velocity
$v_{{c}}$  extracted from the time dependence of the HWHM,
second, the expansion velocity  $v_{{r}}=d \tilde R(t)/dt$, where $\tilde R(t)=\sqrt{R^2(t)-R^2(0)}$, related to the
time dependence of the radius of the cloud, 
\begin{equation}
R^2(t) = \frac{1}{N}\sum_{i}  \langle \hat n_i(t)\rangle (i-i_0)^2d^2\,,
\label{eq:rn}
\end{equation}
 and third, we analyze the average velocity 
\begin{equation}
v_{\mathrm{av}}(t)=\sqrt{\frac{1}{N\hbar^2}\sum_k \left(\frac{\partial \epsilon_k}{\partial k}\right)^2 n_k(t)}\,,
\label{eq:vex}
\end{equation} 
where $\epsilon_k=-2J \cos(kd)$ is the one-particle dispersion.
%Ideally, in the case of a ballistic expansion, $\tilde R=v_{{r}} t$ at all times. This behavior is strictly only realized
%at $U=0$ and $U=\infty$, where in both cases, the expansion is ballistic
%and $v_{{r}}=v_{\mathrm{av}}(t=0)$.
%For $U=0$, $n_k(t)=n_k(t=0)$ is time-independent,
%while in the  case of $U=\infty$, the
%system can be mapped onto free fermions via the Jordan-Wigner transformation. Under this transformation,
%the real-space densities of hard-core bosons and fermions are identical, yet the fermionic
%momentum distribution is time-independent during the expansion, while the bosonic momentum distribution function undergoes a transient dynamics before finally approaching that of the fermions.
%\cite{Rigol2004b,Rigol2005a}
%Yet, consistent treatment of all interaction strengths requires the use of $n_k$ of the hard-core bosons, which undergoes transient dynamics \cite{Rigol2005a,Rigol2004b}).  
 
We further compute the time dependent probability for  multiply-occupied sites (i.e., sites with $\eta_i >1$). 
Numerically, we compute the fraction $\nu_{{h}}$ of atoms on multiply occupied sites
from
\begin{equation}
\nu_{{h}} =  \frac{1}{N}\sum_i \sum_{m=2}^{N_{{b}}}  \, m\,\langle {\hat n}_{m,i}\rangle\,.
\label{eq:nh}
\end{equation}
where ${\hat n}_{m,i}$ measures the probability of finding $m$ bosons on site $i$, $\langle \eta_i | {\hat n}_{m,i} | \eta_i \rangle = \delta_{\eta_i,m}$, where $| \eta_i \rangle$ are the local Fock states on site $i$. 
This should be compared to $f_{{h}}$, the quantity that can be accessed in our experiment. Numerically, we compute it as
\begin{equation}
f_{{h}} = \frac{1}{N} \sum_i (2\langle {\hat n}_{2,i}\rangle +  2\langle {\hat n}_{3,i}\rangle+ 4 \langle {\hat n}_{4,i}\rangle + \dots)\,.
\label{eq:fh}
\end{equation}
We have also studied initial states with a finite density of holes (i.e., some $\eta_i=0$ in Eq.~\eqref{eq:init}),
surrounded by sites with $\eta_i=1$. We find that such single hole defects (in the absence of doubly occupied sites)
do not influence the expansion velocity in the hard-core limit $U=\infty$ and therefore we do not discuss these results here any further.

\subsection{Sudden expansion starting from initial states with exactly one boson per site}

\subsubsection{Time dependence of density profiles}
Figure~\ref{fig:epaps1} shows t-DMRG results for the time dependence of the density $\langle \hat n_i(t)\rangle$ (top row, (a-f)), the  doublon density $\langle \hat n_{2,i}(t)\rangle$
(middle row, (g-l)), and the density of triple occupancy $\langle \hat n_{3,i}(t)\rangle$ (bottom row, (m-r)).
The density profiles at $U/J=2$ and $4$ exhibit a bimodal structure: fast, ballistic tails and a slowly expanding high-density core.  

The density of multiply occupied lattice sites is zero both in the initial state and at all times 
for $U=\infty$. After opening the trap and for  $U/J< \infty$, 
multiply occupied sites are dynamically generated. A net production (resulting in an increase of $\nu_{{h}}$, Eq.~\eqref{eq:nh}) occurs due to initial relaxation dynamics following the quench to finite $U/J$. This has to be contrasted to expansions that start from a system that is in thermal equilibrium (compare, e.g.,  Refs.~\onlinecite{Rigol2004b,Rigol2005a,Schneider2012,Langer2012})
for which a non-trivial time evolution is solely due to the quench of the trapping potential to zero. 
Numerically, we observe that mostly double occupancies appear. 
At $U=0$, multiply occupied sites do not have an effect on the expansion speed, whereas for $U/J\gtrsim4$ isolated doublons tunnel slower than single bosons since their effective hopping matrix element is $\sim J^2/U$ at large $U$. In addition, 
we observe the effect of quantum distillation \cite{Heidrich-Meisner2009} 
at $U/J=20$, where the doublons move towards the center of the cloud and stay there up to the maximum simulation time. In the case of bosons (as compared to fermions), 
there is an attractive interaction between doublons, enhancing the stability of strings of doubly occupied sites over time \cite{Petrosyan2007,Muth2012}. 

\subsubsection{Cloud radius and expansion velocity}

\begin{figure}[tb]
\includegraphics{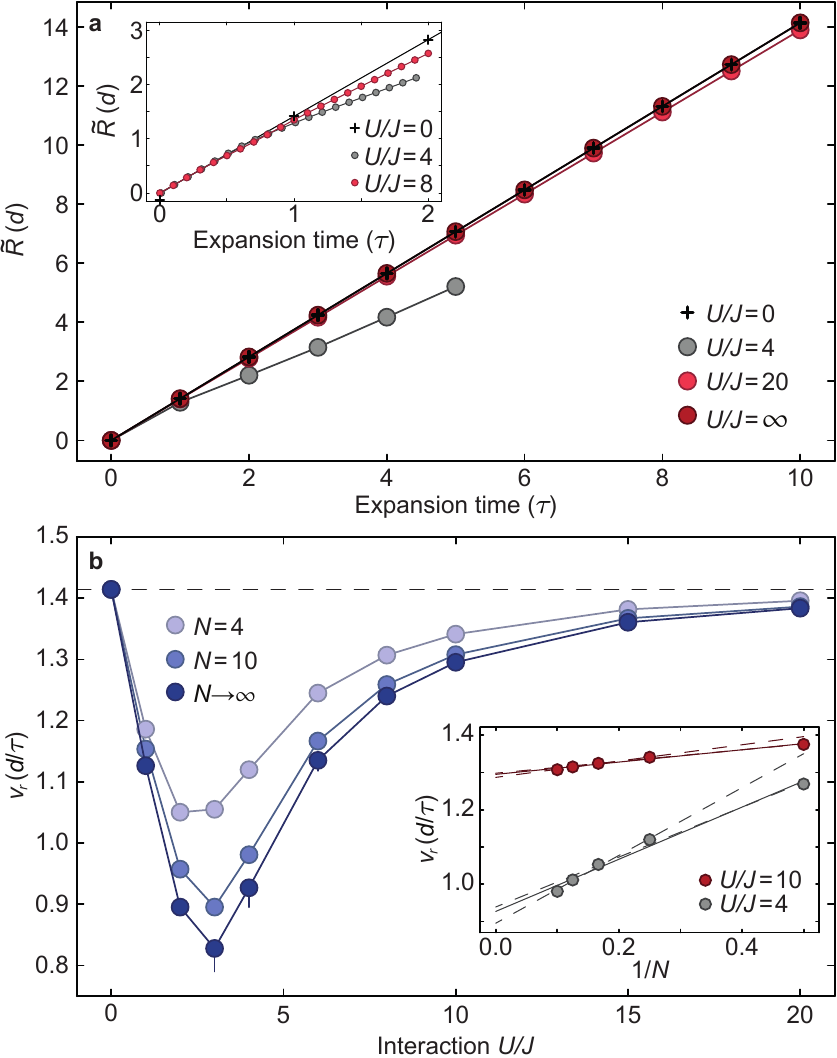}
\caption{{  Time dependence of the radius $\tilde R$ of the expanding cloud}. (a) $\tilde R(t)$ for $U/J=0,4,20,\infty$. Inset: Zoom-in on the transient dynamics  for  $U/J=0,4$ and $8$.
(b) Expansion velocity $v_{{r}}$ for $N=4,10$
and extrapolated in $1/N$ to $N\to \infty$. The dashed line indicates $v_{{r}}=\sqrt{2}\, (d/\tau) $.
Inset: Finite-size extrapolation for $U/J=4$ and $U/J=10$. The solid line uses all data points, the dashed lines exclude either the smallest or largest $N$ 
from the fits. The difference between the fit results of the solid line and the dashed lines determines  the errors bars shown in the main graph. }
\label{fig:epaps2}
\end{figure}

 We analyze  the radius $\tilde R(t)$ and the expansion velocity $v_{{r}}$ from data such as those 
shown in Fig.~\ref{fig:epaps1}. 
Figure~\ref{fig:epaps2}(a) shows the time dependence of the radius $\tilde R(t)$ 
for $N=10$ bosons and $U/J=0,4,20,\infty$. 
In the two exactly solvable limits $U=0$ and $\infty$, the radius increases linearly in time, i.e., $\tilde R(t) =v_{{r}} t$ with $v_{{r}}=v_{\mathrm{av}}(0)$, as expected for these ballistic dynamics. In the $U=\infty$ case, this can be seen using the mapping to spinless fermions, whose quasimomentum distribution is time-independent
 $n^{\mathrm{f}}_{k} (t)=\mathrm{const}$.
 
At intermediate values of $U$, we observe fast transient dynamics for $t\lesssim \tau$:
up to roughly this point in time, the radius increases linearly  with a slope that is independent of $U$. For $t\gtrsim \tau$, it 
crosses over to a linear increase with a {\it smaller} slope, with the slope now depending on $U$ (inset in Fig.~\ref{fig:epaps2}(a)). 
The expansion velocity $v_{{r}}$  shown in Fig.~2 of the main text and in Fig.~\ref{fig:epaps2}(b) corresponds to this 
smaller slope. 
%We  argued in the main text that the  transient behavior is related to the dynamical formation of doublons, which result in a reduction of kinetic energy. 
The deviation from a linear increase
of the radius with a constant velocity is an indication of non-ballistic dynamics for small and intermediate values of $U$.

 In Fig.~\ref{fig:epaps2}(b), we plot $v_{{r}}$ for  finite particle numbers $N=4$ and $N=10$. We also extrapolated 
$v_{{r}}$ to $N\to \infty$ by using data for  $N=2,4,6,8,10$ and a fitting function $v_{{r}}(N)=v_{{r}}+a/N$ (inset). 
We only took into account data for the radius $\tilde R(t)$ up to the same maximum time
for all values of $N$.  We therefore cannot exclude a systematic error due to the limited
timescales that can be reached in t-DMRG simulations (the larger $N$, the shorter the accessible times). Qualitatively,
we tend to overestimate $v_{{r}}(N\to\infty)$ in the vicinity of the minimum of $v_{{r}}$.
The result of this extrapolation is included in Fig.~\ref{fig:epaps2}(b) as well and 
  we conclude that the  pronounced drop in the expansion velocity for $0<U/J<10$  is robust against
finite-size effects. 
 %At $U=0$ limit and in the limit of hard-core bosons, 
%finite-size effects cease to matter and the expansion velocity is $v_{{r}}=v_{\mathrm{av}}(t=0)=\sqrt{2}\,(d/\tau)$ in agreement with Eq.~\eqref{eq:vex}.

We stress that, in our case, the expansion velocity $v_{{r}}$ is not simply a measure of the release energy $E=E_{\mathrm{kin}}+E_{\mathrm{int}}$, which vanishes independently of $U$. This is in contrast to 
previously studied expansions of interacting bosons \cite{Holland1997,Jukic2009} or fermions \cite{Langer2012} from an initial state that is in equilibrium.
For instance, for the Tonks-Girardeau gas or weakly interacting Bose gases in a free space expansion, in the asymptotic limit, $v_{{r}} \propto \sqrt{E}$~\cite{Holland1997,Jukic2009}.

\subsubsection{Dynamical formation of multiply occupied lattice sites}
\label{sec:tdmrg_doublons}
We argue that the fast transient dynamics evident in $\tilde R(t)$ are due to local relaxation processes. 
It is instructive to consider two cases:
 (i) the sudden expansion under some value of $U/J$ realized in the experiment, and (ii)
the time evolution  without opening the trap, but after quenching to a finite value of $U/J<\infty$.
(for theoretical studies on quantum quenches of the interaction strength in the Bose-Hubbard model see, e.g., Refs.~\cite{Schutzhold2006,Kollath2007,Cramer2008a,Roux2009,*Rigol2010,*Roux2010a,Roux2010,Biroli2010,Navez2010}).

\begin{figure}[t]
\includegraphics{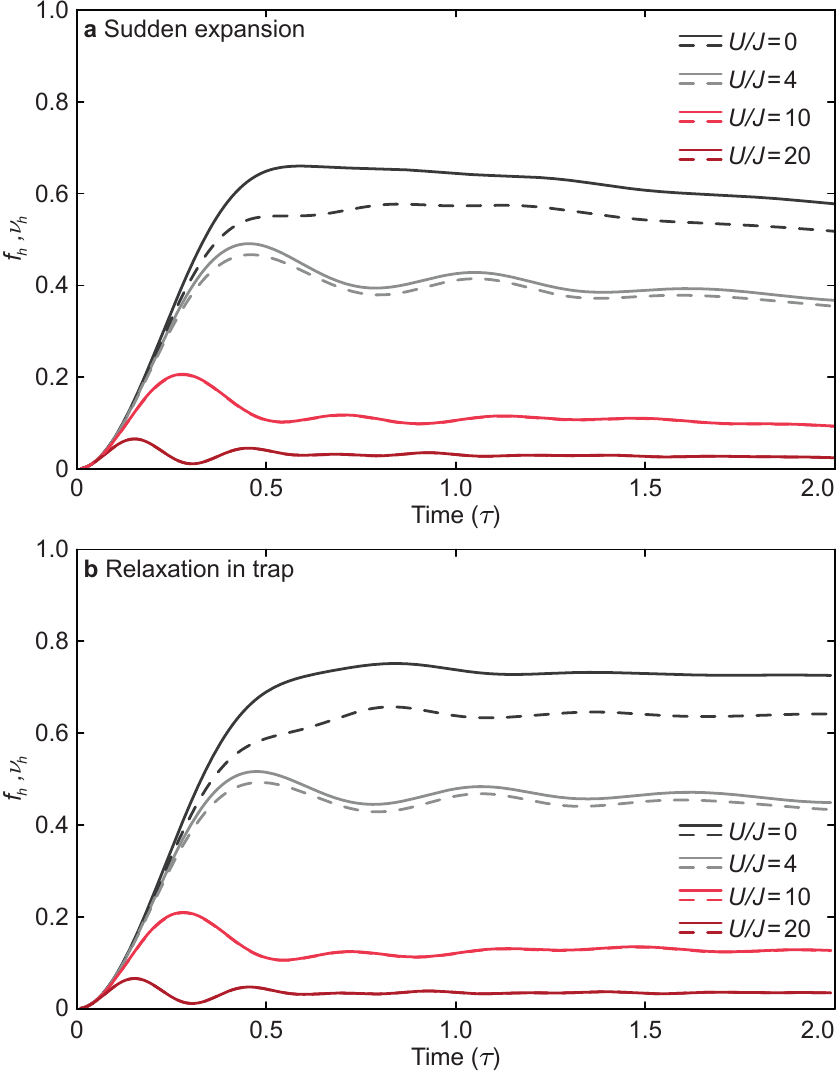}
\caption{{ Fraction  of atoms on multiply occupied sites.}   
 We show both the exact fraction $\nu_{{h}}$ (solid lines) and the 
experimentally accessible measure $f_{{h}}$ (dashed lines), both computed with t-DMRG for the expansion from a state with $\eta_i=1$ (Eq.~\eqref{eq:init}), for $U/J=0,4,10,20$, $N=10$, and $N_{{b}}=N$. 
(a) Evolution during the expansion.
(b) Evolution after a quench in $U/J$, without opening the trap. 
These data suggest that the transient time associated with the formation of higher occupancies is given by $t\sim 0.5\,\tau$ such that this initial relaxation is purely local.}
\label{fig:epaps3}
\end{figure}

 Figure~\ref{fig:epaps3}(a) and (b)  show the time dependence of the fraction of  bosons  on multiply occupied sites for $U/J=0,4,10,20$ for cases (i) and (ii), respectively.  
We present both the total higher occupancy $\nu_{{h}}$ (solid lines) and the experimentally
accessible quantity $f_{{h}}$ (dashed lines).
For scenario (ii),  the initial  state Eq.~\eqref{eq:init} with $\eta_i=1$ is not an eigenstate except for $U=\infty$, and therefore
the system explores phase space, resulting in a  dynamical formation of multiply occupied lattice sites, i.e. $\nu_{{h}}>0$.
The fraction of atoms on multiply occupied sites  is similar in  cases (i) and (ii),
corroborating the notion  that the net production of higher occupancies is a local process, and therefore not a consequence of the expansion as such.

Furthermore, the figure demonstrates that, for any $U/J<\infty$, the system forms higher occupancies 
on a timescale of $t\approx 0.5\tau$.  Both $\nu_{{h}}$ and $f_{{h}}$  saturate at $U$-dependent values  and, in the expanding case, slowly decay at larger times with small oscillations. 
Moreover, for $U/J>4$, $\nu_{{h}}\approx f_{{h}}$, indicating that in this regime only double occupancies are formed while higher occupancies are suppressed.
Our  t-DMRG results for $f_{{h}}$ are in qualitative agreement with the experimental data presented in Fig.~2(b) of the main text concerning the
timescales of the formation and the decay of higher occupancies.
We ascribe quantitative differences to the presence of hole defects in the experiment and the inhomogeneity due to the harmonic
trap in the experiment.
 
 \subsubsection{Time dependence of the quasimomentum distribution}
%In the main text, we argue that the 
%decrease of the expansion velocity at large $U/J$ is a consequence of the dynamical formation of higher occupancies (mostly doublons).
%In the small $U$ regime, doublons (and larger objects) cannot be thought of as separate entities.
%To understand the decrease of the expansion velocity as $U$ increases away from $U=0$,
%it is instructive to study the time dependence of the momentum distribution $n_k$. 

\begin{figure}[t]
\includegraphics{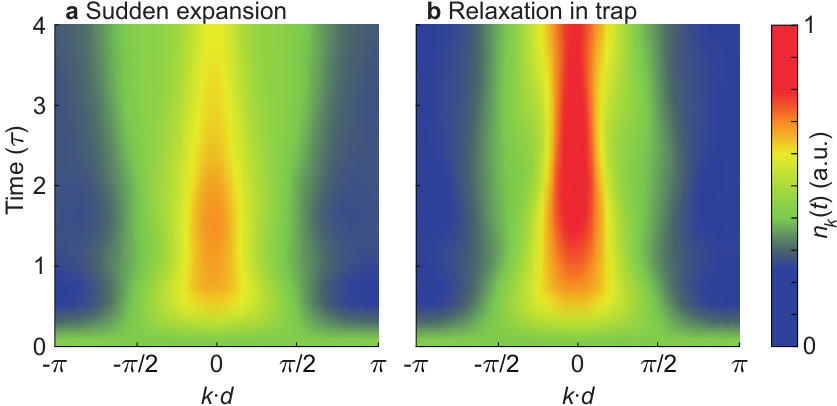}
\caption{{ Time evolution of the quasimomentum distribution $n_k(t)$.} (a) Sudden expansion. (b) Relaxation without opening the trap. Both dynamics start from the state given in Eq.~\eqref{eq:init} with $\eta_i=1$ at $U/J=1$ for $N=10$ using $N_{{b}}=N$.} 
\label{fig:epaps4}
\end{figure}

The fast transient dynamics as opposed to the slower dynamics during the expansion can further be elucidated by
considering the time dependence of the quasimomentum distribution $n_k$.
In Fig.~\ref{fig:epaps4}(a) and (b), we display t-DMRG results for $n_k(t)$ for the two cases (i) and (ii), respectively,  with $U/J=1$. 
In both scenarios, $n_k(t)$ develops a maximum at $k=0$ on transient timescales $t\sim\tau$. This maximum remains stable in case (ii), 
while in case (i), the central peak slowly dissolves at later times during the expansion. 
 
\begin{figure}[t]
\includegraphics{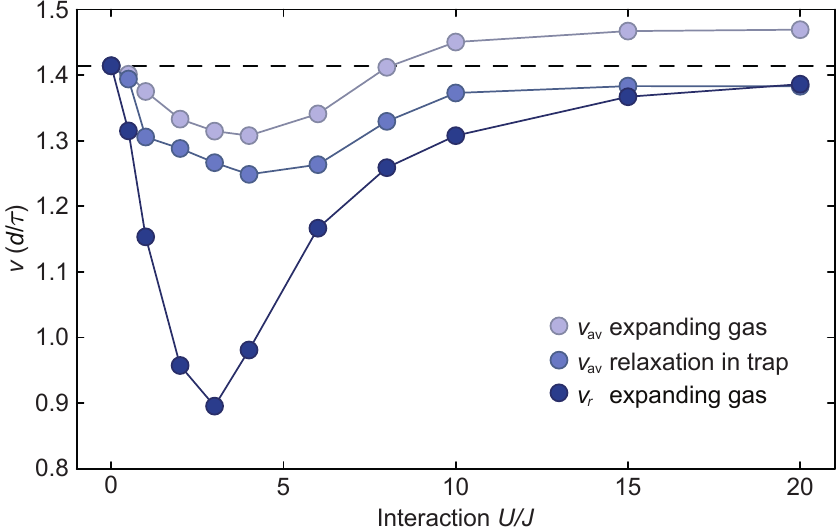}
\caption{{Expansion velocity.} Average expansion velocity $v_{\mathrm{av}}$ as a function of $U/J$ for the expanding gas and without opening the trap, both at time $t=\tau$.
We compare this to $v_{{r}}$. The dashed line indicates $v=\sqrt{2}\,(d/\tau)$. All data are calculated for $N=10$ and $v_{\mathrm{av}}$ is calculated using $N_{{b}}=N$ in both cases.}
\label{fig:epaps5}
\end{figure}

From such data for $n_k(t)$ and for various values of $U$, we calculate the average expansion velocity $v_{\mathrm{av}}$ at time $t=\tau$ as a function of $U/J$ and compare it to $v_{{r}}$ (Fig.~\ref{fig:epaps5}). 
Indeed, $v_{\mathrm{av}}(t=\tau)$ of the expanding gas has a weak minimum at $U/J\sim 3$ and 
increases for larger $U/J$, eventually exceeding $\sqrt{2}\,(d/\tau)$.  The latter, $v_{\mathrm{av}}>\sqrt{2}\,(d/\tau) $, occurs  because of  the dynamical quasi-condensation~\cite{[The dynamical emergence of coherence was also observed for fermions
in: ]Heidrich-Meisner2008} at large $U$ where predominantly the quasimomenta at
$k=\pm \pi/(2d)$ become occupied \cite{Rigol2004b,Rodriguez2006a}. 
The minimum of $v_{\mathrm{av}}$ is present both for the actual expansion and
the relaxation of the trapped gas without opening the trap. 
Therefore, we conclude that the decrease of the expansion velocity, measured through either $v_{{c}}$ or $v_{{r}}$, is partially due to
the relaxation dynamics of the quasimomentum distribution at short times with a tendency of occupying small momenta with a
small velocity $v_k\ll 2(d/\tau)$ ($v_k=(1/\hbar) \partial\epsilon_k/\partial k=(2d/\tau)\sin{(kd)}$). However, as Fig.~\ref{fig:epaps5} clearly shows, $v_{{r}}\ll v_{\mathrm{av}}(t=\tau)$ at $U/J\lesssim 10$,  
indicating that interactions {\it during} the expansion lead to an additional substantial drop of $v_{{r}}$ and $v_{{c}}$.

\begin{figure}[t]
\includegraphics{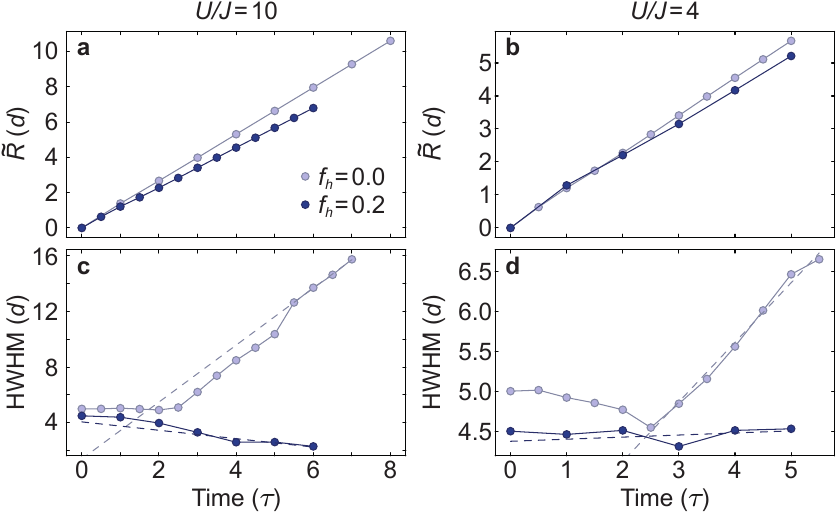}
\caption{{ Sudden expansion in the presence of doublons in the initial state.} 
(a), (b) Radius $\tilde R(t)$  for $U/J=10$ and $U/J=4$. (c), (d) Time dependence of the HWHM for 
$U/J=10$ and $U/J=4$. The data for $f_{{h}}\not= 0$ were obtained by averaging over
all distributions of doublons ($\eta_i=2$) while keeping all other occupied sites in the initial state at $\eta_i=1$. Dashed lines are linear fits to the last four data points that are used to obtain $v_{{c}}$ 
($N=10$ in all simulations).}
\label{fig:epaps6}
\end{figure}

\subsection{Expansion in the presence of doublons in the initial state}
In this section we investigate initial states that have
a finite concentration of sites with an occupancy of $\eta_i=2$. 
Fixing the particle number to $N=10$, we generate all possible realizations with such defects for a given $f_{{h}}\not=0$ and calculate the averaged time dependent density. 

Figure~\ref{fig:epaps6} shows our results for the radius $\tilde R(t)$ and the HWHM at $U/J=10$ and $U/J=4$, 
comparing the expansion from an initial  state with $\eta_i=1$ only ($f_{{h}}=0$) to the ones with defects ($f_{{h}}\not= 0$). Already 
for the clean state there is a transient behavior in the HWHM before 
a linear increase in time sets in, which is consistent with the experimental observations 
discussed in Sec.~\ref{Exp_det_vc} (compare Fig.~\ref{fig:Dipole_and_lin_fits}(c)). A similar behavior of the HWHM was also seen in a theoretical study of the sudden expansion of a 
Tonks-Girardeau gas \cite{Campo2006}. 
 
Upon adding doublons to the initial state, the slope of the radius $\tilde R(t)$ decreases, as  shown in 
Fig.~\ref{fig:epaps6}(a) for $U/J=10$. The effect is small since $\tilde R(t)$ is dominated by the fast moving ballistic tails, which are unaffected by the presence of a few doublons (see Fig.~\ref{fig:epaps1}). 
The HWHM is, however, much more sensitive to the presence of
doublons in the initial state: already for $f_{{h}}=0.2$, its slope, $v_{{c}}$,  is zero or slightly negative. 
We observe this dramatic dependence of $v_{{c}}$ on $f_{{h}}$ for both $U/J=4$ and $U/J=10$.
These numerical results agree well with the experimental data for $v_{{c}}=v_{{c}}(f_{{h}})$ shown in Fig.~4 of the main
text.

At smaller $U/J\sim 1$, our t-DMRG results do not show any strong effect of the presence of 
doublons in the initial state on either $v_{{c}}$ or $v_{{r}}$, in contrast to the experimental results.
 We attribute this deviation to the different particle numbers ($N\sim 10$ in t-DMRG simulations versus
$N\sim 80$ per tube in the experiment).

\end{document}